# HiddenLens: A Multi-Layer LLM Framework for Detecting and Resolving Hidden Assumptions in Software Requirements

**Tanu Singh[1], Aashna Suman[2], Vinod Patidar[3*]**

School of Computer Science, UPES, Dehradun, Uttarakhand, India

[1]tanu.singh@ddn.upes.ac.in, [2]aashna.15692@stu.upes.ac.in

[3*]vinod.patidar@ddn.upes.ac.in (Corresponding Author)

## ABSTRACT

Software requirements written in natural language often omit important assumptions about expected performance, operating conditions, user behaviour, and domain-specific constraints. This gap between explicitly stated requirements and stakeholders' implicit assumptions can influence architectural choices, testing approaches, and deployment configurations. If these assumptions are not identified until implementation, correcting the resulting problems can be far more costly than addressing them at the early stages of requirements engineering. Although ambiguity and vagueness in requirements have been widely studied, less attention has been given to uncovering assumptions embedded in requirements that otherwise appear clear and reasonable. To address this gap, this paper proposes **HiddenLens**, a novel six-layer BERT-LLM pipeline and the first framework engineered specifically to detect, classify, question, and resolve hidden assumptions in software requirements, rather than merely flagging the existence of ambiguity. Layer 1 applies BERT attention entropy to localise the specific spans of a requirement carrying the most unresolved linguistic weight. Layer 2 passes these spans to an LLM using CoT prompting, which reasons step-by-step to articulate the precise hidden assumption each span conceals and the failure risk if that assumption is wrong. Layer 3 classifies each detected assumption with a fine-tuned BERT classifier across six non-functional requirement categories: performance, security, scalability, usability, reliability, and domain. Layer 4 generates context-aware, ranked clarification options for the analyst via FS LLM prompting, turning each assumption into concrete, answerable choices. Layer 5 is a human-in-the-loop validation step, where the analyst confirms, edits or requests a layman-mode rephrasing of each clarification before anything is finalised. Layer 6 closes the loop by rewriting the original requirement, substituting the analyst-confirmed resolution in place of the ambiguous term, and producing a requirement that is explicit, testable, and traceable back to the assumption it resolves. This paper also introduces the **HiddenLens Corpus (HLC)**, a dedicated dataset comprising 312 requirements annotated with 847 hidden assumptions across six NFR categories. Inspired by the PROMISE and PURE datasets, HLC addresses the lack of annotated instances for hidden-assumption detection in the existing corpora. To the best of our knowledge, it is the first dataset specifically developed for this task. HiddenLens outperforms the keyword-based baseline by 71% and the QuARS-Adapted baseline by 61%, achieving an end-to-end F1-score of 0.79, while its BERT-based NFR classifier achieves an F1-score of 0.847. These findings indicate that HiddenLens can help analysts identify and resolve hidden assumptions during requirements engineering before they propagate to later stages of software development.

**Keywords:** Requirements Engineering, Hidden Assumptions, Large Language Models, BERT, Prompt Engineering, NFR Classification, Chain-of-Thought, Requirements Quality

# 1. Introduction

Requirements Engineering (RE) is the process of capturing, documenting, and agreeing on what a system must do before a single line of code is written, and it is widely recognized as the most consequential stage of the software development lifecycle (SDLC). The single highest-leverage activity in software development is ensuring that the requirements are correct. Previous studies show that defects identified during the requirements stage can cost up to 100 times less to fix than those discovered at later stages (Boehm, 1984; Femmer et al., 2017). RE shapes every downstream decision, system architecture, test coverage, resource allocation, and ultimately whether the delivered product matches stakeholder requirements. Basically, RE is the blueprint stage of the SDLC. If software is built on poorly stated requirements, the project will have issues at every layer and lead to software failures. RE has long been recognised for its importance but still proves to be a very underserved and error-prone stage in practice. Such issues lead to constant defects that eventually turn into project failures.

The consequences of such failures have been examined extensively in previous studies. One such work has identified requirements defect in the top three root causes of systemic failures. The rest two include poor project management and inadequate testing (Johnson, 2020; Leffingwell & Widrig, 2003). There are three major types of requirements problems that can lead to failures. First, incompleteness, wherein required behaviours are not captured. Second, inconsistency, where requirements contradict each other across documents. Third, ambiguity, where the language used allows multiple interpretations. Ambiguity is one of the most pervasive and difficult-to-detect requirements problems.

Ambiguity does not appear out of nowhere; it sneaks in through the way requirements are written, and almost always, that means the use of natural language. Stakeholders use plain language because it is easy for everyone involved to understand, whether they are developers, managers, or clients. It is a genuine strength of natural language, but it comes with a hidden cost. When stakeholders agree that "*the system shall respond quickly*" or "*data shall be stored securely*" everyone appears in agreement but having different interpretations of these statements. To one person, "*quickly*" may mean under two seconds; however, for another person it may mean instantly. Similarly, "*securely*" might mean full encryption to a developer, but for a business manager it might mean that the system does not get hacked.

**Table 1: Examples: Requirements and Vague Words**

| Requirements | Vague Words |
|---|---|
| The system shall process user requests quickly. | Quickly |
| Patient data shall be stored securely. | Securely |
| Nurses can update patient status in real time. | real time |
| The system should respond immediately in emergencies. | Immediately |
| User data shall be backed up regularly. | regularly |
| The interface shall be intuitive for new users. | intuitive |
| The application shall be user-friendly. | user-friendly |
| The system shall support a large number of concurrent users. | large number |
| Reports shall be generated efficiently. | efficiently |
| The system shall be highly available. | highly available |
| Administrators shall be able to easily configure system settings. | easily |
| The system shall handle errors gracefully. | gracefully |

Additional examples of requirements and vague words are listed in Table 1. In such a scenario, no one is wrong exactly, but stakeholders are not sharing the same understanding of requirements. These are not spelling mistakes or formatting issues; rather, they are gaps in meaning hidden behind words that look perfectly reasonable on paper. The real problem is that these gaps remain unnoticed during requirements review, because the requirement appears to be complete and well written. The intended meaning of the word "*quickly*" might not be questioned until the system is built and the client finds that the delivered system does not meet their expectations.

As these vague words are rarely stated explicitly or questioned during the requirements review process, stakeholders often have to fill in these gaps with their own interpretations that become hidden assumptions. Such assumptions may appear harmless at the early stages of development; however, they can lead to different expectations among stakeholders and consequently the deviations between the specified and expected system behaviour. Identifying these hidden assumptions is therefore important for detecting potential requirements failure at an early stage. Figure 1 illustrates how such hidden assumptions can remain unnoticed during requirements specification and eventually contribute to project failure.

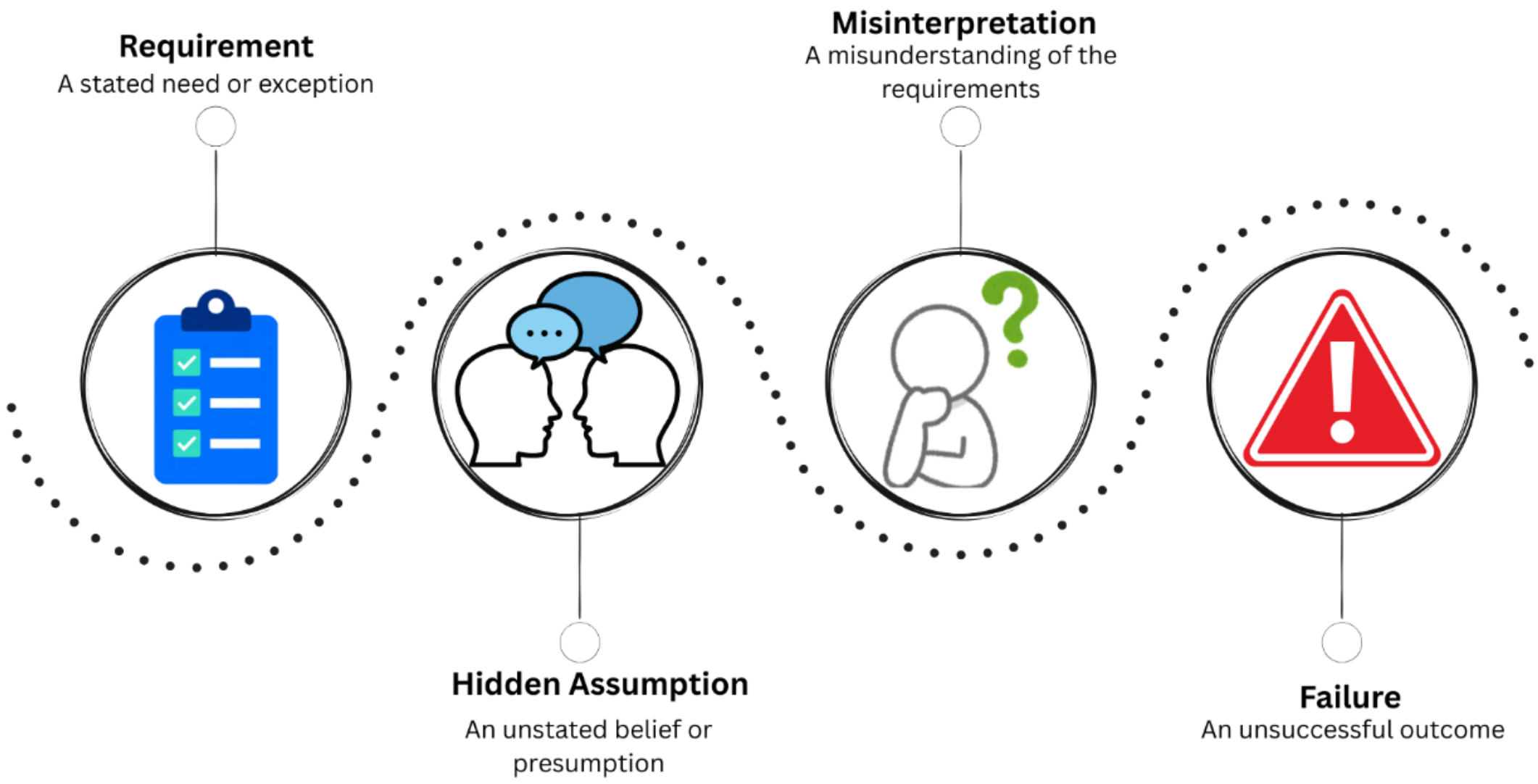

Figure 1. Hidden Assumptions Leading to Project Failure

The research community has not overlooked the problem of hidden assumptions; however, it has largely addressed a different manifestation of the issue. Existing studies on the detection of syntactic and semantic ambiguities (Ferrari et al., 2024; Gervasi et al., 2019), the classification of requirements smells (Femmer et al., 2017), and automated quality assessment against standards such as ISO/IEC/IEEE 29148 (2018); Lubos et al., 2024 mainly focus on poorly specified or incorrectly written requirements. Similarly, tools such as RETA (Bajceta et al., 2022) and QuARS (Lami et al. 2005) can identify ambiguous terms and implicit pronoun references, while rule-based keyword detectors can flag hedging or imprecise language. However, these approaches cannot explicitly identify the underlying assumption obscured by vague terms. Consequently, they detect symptoms of incompleteness or ambiguity but cannot identify the specific information missing.

To bridge this gap, this paper presents **HiddenLens**: a novel six-layer Bidirectional Encoder Representations from Transformers- Large Language Model (BERT–LLM) pipeline for detecting, classifying and resolving hidden assumptions in software requirements. Rather than being another ambiguity detector, HiddenLens addresses a fundamentally different question: what does a requirement silently depend on? Instead of merely identifying what is ambiguous in a requirement, HiddenLens seeks to uncover the underlying assumptions, determine their nature, facilitate their validation and resolve them by formulating requirements that are clearer and more explicit.

The proposed framework follows an end-to-end process that combines automated linguistic analysis, LLM-based assumption identification & clarification generation, requirement classification, and human-in-the-loop (analyst) validation, culminating in the rewriting of the original requirement. This design enables HiddenLens to move beyond detecting surface-level vagueness to identifying and resolving assumptions that may remain implicit in otherwise seemingly acceptable requirements. The framework is evaluated using requirements drawn from the publicly available PROMISE (Cleland-Huang et al., 2007) and PURE (Ferrari et al., 2017) datasets, supplemented by industrial documents from healthcare partners and synthetically generated examples, with the resulting dataset manually annotated for hidden assumptions across six non-functional requirement (NFR) categories. We refer to this dataset as the ***HiddenLens Corpus (HLC)*** and to support reproducibility, we publicly release its synthetic portion: 312 requirements with 847 annotated hidden assumptions as a companion resource to PROMISE and PURE. Through this integrated approach, HiddenLens aims to support requirements analysts and software developers in improving the clarity, completeness, and interpretability of requirements at an early stage of software development.

The major contributions of this paper are as follows:

- **Novel End-to-End Framework:** HiddenLens is a six-layer pipeline that integrates BERT-based attention entropy, LLM-based chain-of-thought reasoning, fine-tuned NFR classification, context-aware suggestion generation, analyst validation through a human-in-the-loop mechanism, and automated requirement rewriting. It is the first end-to-end framework designed to detect and resolve hidden assumptions in software requirements.
- **Prompt Engineering Strategy:** The study provides an in-depth comparison of zero-shot (ZS), few-shot (FS) (Brown et al., 2020), and chain-of-thought (CoT) prompting (Wei et al., 2022) for hidden-assumption detection and clarification generation. Based on this comparison, an appropriate prompting strategy is selected for each task.
- **Annotated Dataset:** The HiddenLens Corpus (HLC) comprises 312 requirements containing 847 annotated assumptions across six NFR categories and is made available to the research community.
- **Empirical Evaluation:** The BERT classifier obtains an F1-Score = 0.847 and the human evaluation of hidden assumption resolution yields precision = 0.81, recall = 0.78 and F1-score= 0.79. Therefore, the proposed approach achieves substantive improvements over rule-based (Fabbrini et al. (2001)) and keyword-based (Wilson et al. (1997)) baselines. It is further supported by a detailed error analysis and an ablation study of various prompt strategies.

The rest of the paper is organized as follows. Section 2 reviews related work on requirements ambiguity detection, NFR classification, and LLM applications in requirements engineering, along with prior approaches that flag ambiguity but do not resolve it. Section 3 formally defines a hidden assumption and introduces the six-category taxonomy used throughout the paper. Section 4 presents the six-layer HiddenLens framework in detail, from BERT attention entropy localisation and chain-of-thought assumption detection through NFR classification, clarification generation, analyst validation, and automated requirement rewriting. Section 5 describes the construction of the HiddenLens Corpus (HLC) and the annotation process behind it. Section 6 details the experimental setup, including model configuration, baseline methods, and both automated and human evaluation metrics. Section 7 presents the quantitative results, a comparative analysis against prior work, and sample outputs. Section 8 walks through a case study on a Hospital Patient Management System to illustrate the framework end-to-end. Section 9 discusses threats to validity, and Section 10 concludes with a summary of contributions and directions for future work.

## 2. Related Work

Early work on ***pattern-based detection*** emerged in the 1990s. One of the earliest systems, the Automated Requirements Measurement (ARM) tool, examined requirements specifications for linguistic indicators of poor quality, including vague, weak, optional, and subjective expressions (Wilson et al., 1997).In parallel, Ambriola and Gervasi (1997) explored the automated processing of natural-language requirements. These efforts helped make linguistic defects amenable to automated analysis. QuARS (Lami et al. 2005) subsequently extended this direction by combining lexical dictionaries with morphological and syntactic analysis to detect ambiguity, understandability, and specification-completeness problems (Fabbrini et al., 2001; Lami et al., 2004).

The work on ***conceptual and heuristic refinement*** shifted the attention from patterns to the thinking behind them. Berry et al. (2003) made an important point that not every ambiguity is similar. Some of it is deliberate, a real expression of uncertainty on the part of stakeholders, and some of it is accidental, just a consequence of bad writing. They argued that automated tools should be able to differentiate these, a nuance that most tools overlooked. This line of thought was followed and formalised by Femmer et al. (2017), who introduced the notion of requirements smells: named as anti-patterns, similar to code smells, that indicate a degradation in quality. Their Smella tool used part-of-speech tagging and heuristic rules to find these patterns at scale, giving practitioners a common vocabulary for requirements defects. However, it could only report that a smell was present and could not diagnose what was missing, like its predecessors.

The advancements in ***statistical machine learning*** subsequently expanded the range of features detectable beyond those captured by manually formulated rules. Winkler and Vogelsang (2016) applied convolutional neural network (CNN) and long short-term memory (LSTM) architectures to the PROMISE NFR dataset (Cleland-Huang et al., 2007), learning distributed representations of requirements text rather than relying on bag-of-words features and thereby improving non-functional requirement classification. However, their models still required substantial amounts of labelled data for each category, could not be transferred across domains without retraining, and were limited to classification. For example, they could classify a requirement as performance-related but could not identify its implicit assumptions regarding performance. Arora et al. (2016) trained classifiers using annotated corpora of industrial requirements and substantially outperformed preceding syntactic approaches. Nevertheless, these classifiers faced the same limitation: a defect had to be explicitly discernible in the text. Similarly, Guo et al. (2017) applied a deep-learning architecture combining word embeddings with bidirectional gated recurrent units (GRUs) to software traceability, substantially outperforming vector-space and latent semantic indexing baselines. Across both tasks, the richer neural representations improved on earlier surface-level methods, but the object of analysis remained content already present in the text, not the information it silently left out. Abad et al. (2017) compared topic-modelling and Naive Bayes (NB) approaches on the same repository and showed that preprocessing choices alone could substantially shift classifier performance, underscoring how sensitive these statistical methods were to pipeline design rather than to the underlying requirements content.

***Transformer-based models*** further advanced requirements classification. Hey et al. (2020) introduced Non-functional and functional Requirements classification using BERT(NoRBERT). This BERT-based model automatically classifies functional and non-functional requirements and achieved a precision of 0.92 on the PROMISE dataset (Cleland-Huang et al., 2007). Later, Luo et al. (2022) proposed Prompt Learning for Requirement Classification (PRCBERT), which applies flexible prompt templates to BERT-style models to achieve a 72.96% F1-score in zero-shot classification on held-out data. In parallel, Ferrari et al. (2024) and Beqiri et al. (2024) studied defect detection in industrial and safety-critical domains, namely automotive & railways, and both have cited domain dependence as the ceiling they could not break through. Lubos et al. (2024) compared LLM-based quality assessment against ISO/IEC/IEEE 29148 (2018) standards and obtained a high recall but modest precision, thus recommending a mixed human-LLM workflow.

A broader perspective emerged with the ***meta-analysis*** conducted by Necula et al. (2024) wherein a systematic review examined 137 Natural Language Processing (NLP)-in-RE studies published in the period 1991 to 2024. They found a pattern that cuts across every generation of tooling in the field consistently targeted what is written poorly, not what is left out entirely. Over time, more complex models, richer libraries of patterns, and deeper representations of the linguistic data were developed, addressing the same class of problems. Comparatively less research has been done on the implicit requirements and the underlying presumptions.

The emergence of ***large language model-based (LLM-based) reasoning*** has expanded the range of questions that can be investigated and answered in requirements engineering (RE). In contrast to pattern-matching approaches, such as rule-based models or fine-tuned classifiers, LLMs can apply domain knowledge, perform reasoning, and generate human-readable explanations. Brown et al. (2020) first showed that LLMs could perform new tasks from only a handful of in-context examples, without fine-tuning, establishing Few-shot (FS) prompting as a viable alternative to task-specific training. Wei et al. (2022) revealed that using the chain-of-thought (CoT) prompts leads to substantial gains in solving inferential problems. Hymel and Johnson (2025) revealed that the GPT-4 achieved expert-level requirements quality with substantially lower costs. In an extensive review, Zadenoori et al. (2025) analyzed 74 works and identified a clear shift from pattern matching to semantic reasoning. Almonte et al. (2025) demonstrated that LLMs can generate plausible non-functional requirements directly from stated functional requirements, resulting strong alignment with expert judgment. This generative task still operated on the information explicitly stated in the original requirements rather than uncovering implicit assumptions that requirements silently rely on. The findings by Unterbusch and Vogelsang (2026) proved that a FS and CoT prompting, combined with persona augmentation, produces the best performance results for requirements benchmarking, adapting to a new domain with very few examples.

Across all the aforementioned technological advances, existing approaches have not directly addressed the identification and resolution of hidden assumptions. LLMs have primarily been employed in requirements engineering to classify, formalise, or rephrase requirements. However, these processes continue to operate on information explicitly present in the text. An LLM may identify a vague term as a requirements smell but does not necessarily articulate the underlying assumption associated with it. The closest antecedent is the analogy-based reasoning framework proposed by Onyeka (2013), which preceded LLMs and was not scalable. As described by Nuseibeh et al. (2001), unvalidated assumptions can translate into architectural flaws. Although the IEC 61508 standard (Häring, 2021) requires such assumptions to be documented, industry reports indicate that these requirements are not consistently followed in practice. Thus, the problem of hidden assumptions has been recognised, and its consequences have been documented in the form of system failures; however, a scalable tool for systematically detecting and addressing these assumptions has remained unavailable.

In order to address each of the aforementioned issues, this study proposes **HiddenLens**, a novel six-layer BERT-LLM pipeline for detecting, classifying, and resolving hidden assumptions in software requirements. The proposed framework is designed to address a gap that has remained unresolved despite decades of tooling in the RE domain. The problem definition adopted in this study is outlined in the next section.

# 3. Problem Definition

This section formulates the problem in two parts. First, it defines a hidden assumption and distinguishes it from a requirement that is merely vague or unclear, as these constitute separate problems addressed by existing tools. Second, it describes the six categories of assumptions considered throughout this paper.

## 3.1 Formal Definition

Let $R$ denote a natural language software requirement sentence. Let $W(R)$ denote the set of all propositions that a competent software architect would need to assume to be true or satisfy to correctly implement $R$. Let $E(R) \subseteq W(R)$ denote the set of propositions explicitly stated in $R$. We define the hidden assumption set $A(R)$ as:

$$A(R) = W(R) \setminus E(R)$$

That is, a hidden assumption $a \in A(R)$ is a proposition that is necessary for the correct implementation of $R$ but is not stated in $R$. This definition distinguishes hidden assumptions from (i) ambiguity, where $a \in E(R)$ but admits multiple conflicting interpretations, and (ii) vagueness, where $a \in E(R)$ but is insufficiently precise. Hidden assumptions are absent, not merely ambiguous or imprecise. This is depicted clearly in Figure 2, which showcases the relationship between requirements and their associated assumptions, sourced from our own analysis.

Hidden assumptions are difficult to detect for three principal reasons. First, they are invisible in the surface text: no linguistic signal marks their absence. Second, they are often domain-dependent: They are often domain-dependent: what counts as a necessary unstated assumption depends on the application domain, deployment context, and stakeholder community. Third, they are spread across multiple NFR categories at once: a single requirement may embed performance, security and environmental assumptions simultaneously.

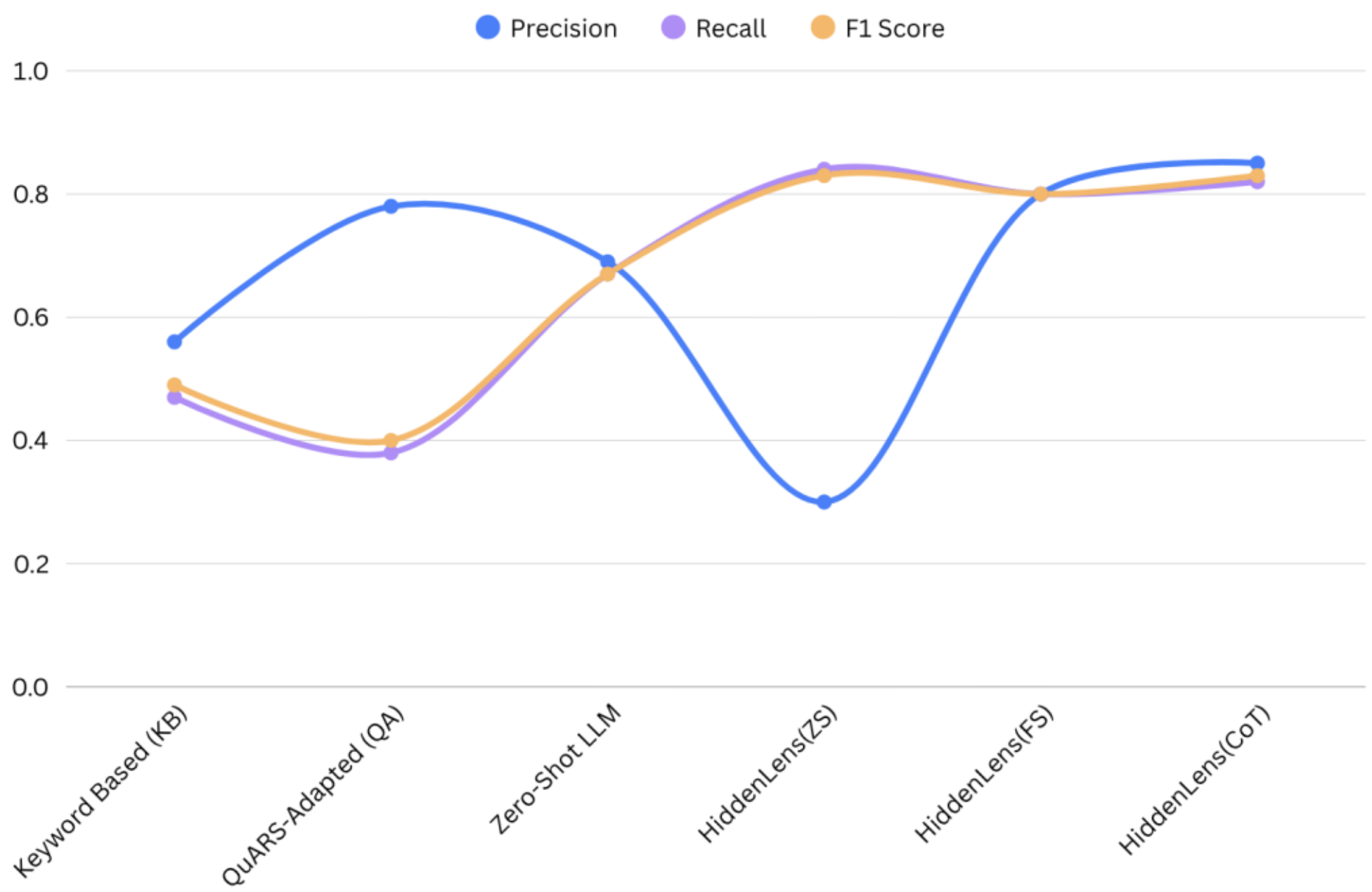

Figure 2. Requirement-Assumption Relationship

### 3.2 Assumption Taxonomy

An analysis of the 312 requirements in the **HiddenLens Corpus (HLC)** and the literature on NFR classification led to the identification of six principal categories of assumptions, which are described in detail in Section 5. Table 2 presents this taxonomy, together with definitions and illustrative examples.

**Table 2: Hidden Assumption Taxonomy**

| Category | Description | Example Assumption |
|---|---|---|
| Performance | Unstated quantitative thresholds for response time, throughput, or resource utilisation. | "Quickly" implies response latency < 200 ms under 1,000 concurrent users. |
| Security | Unstated encryption standards, access control models, or compliance requirements. | "Securely" implies AES-256 encryption and RBAC with principle of least privilege. |
| Scalability | Unstated user load, data volume, or growth trajectory assumptions. | "Large number of users" implies horizontal scaling to 10,000 concurrent sessions. |
| Usability | Unstated user expertise levels, device types, or interface accessibility requirements. | "Easy to use" assumes non-technical administrative staff as primary users. |
| Reliability | Unstated availability SLAs, fault tolerance models, or recovery time objectives. | "Always available" assumes 99.9% uptime with sub-hour RTO. |
| Domain | Unstated domain-specific rules, regulatory mandates, or integration constraints. | "Patient data" implies HIPAA compliance and audit trail requirements. |

## 4. The HiddenLens Framework

HiddenLens is a novel six-layer BERT-LLM pipeline for detecting, classifying, and resolving hidden assumptions in software requirements. It engages an analyst to clarify hidden assumptions and rewrite the requirement to make them explicit. Figure 3 illustrates the full architecture of the proposed HiddenLens framework.

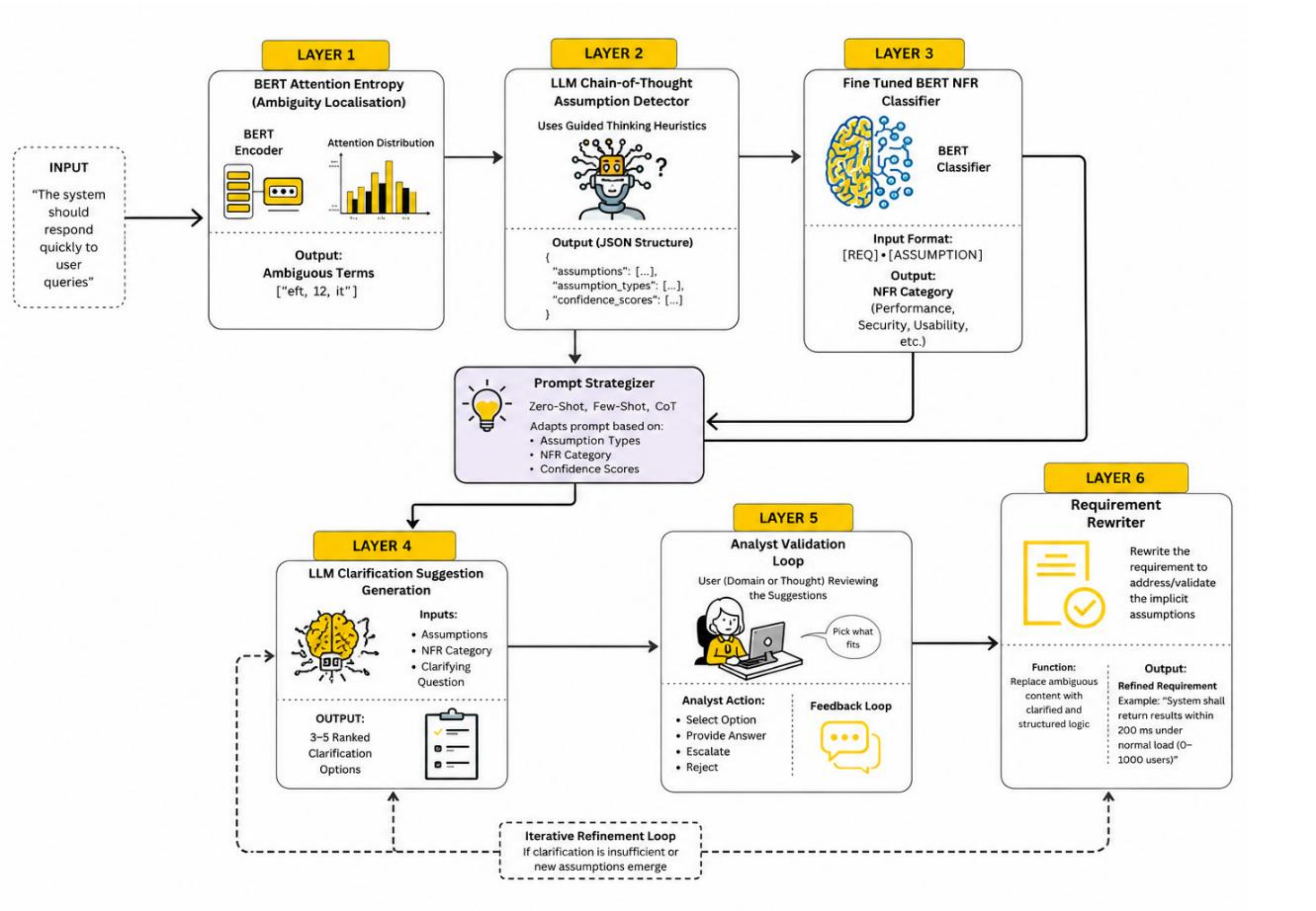

Figure 3. HiddenLens Architecture

Layer 1 applies BERT attention entropy to localise the specific spans of a requirement carrying the most unresolved linguistic weight. Layer 2 passes these spans to an LLM using CoT prompting, which reasons step-by-step to articulate the precise hidden assumption each span conceals and the failure risk if that assumption is wrong. Layer 3 classifies each detected assumption with a fine-tuned BERT classifier across six non-functional requirement categories: performance, security, scalability, usability, reliability, and domain. Layer 4 generates context-aware, ranked clarification options for the analyst via FS LLM prompting, turning each assumption into concrete, answerable choices. Layer 5 is a human-in-the-loop validation step, where the analyst confirms, edits or requests a layman-mode rephrasing of each clarification before anything is finalised. Layer 6 closes the loop by rewriting the original requirement, substituting the analyst-confirmed resolution in place of the ambiguous term, and producing a requirement that is explicit, testable, and traceable back to the assumption it resolves.

In the following subsections, we provide details of each layer of the HiddenLens framework, including its architecture, functionality, and how it helps in identifying and resolving hidden assumptions in software requirements.

### 4.1 Layer 1: BERT Attention Entropy (Ambiguity Localisation)

The pipeline begins by identifying candidate ambiguous terms within the requirement sentence. We exploit the observation that BERT's attention weights, when averaged across heads and layers, are distributed more uniformly across tokens when a term is semantically underspecified; high entropy signals uncertain contextualisation. For each non-stop-word token $w$ in the requirement sentence, we compute:

$$H(w) = -\Sigma\, \alpha_{ij} log(\alpha_{ij}),\ i \in layers, j \in heads,$$

where $\alpha_{ij}$ denotes the mean attention weight assigned t.$w$ by head $j$ of layer $i$. Tokens whose entropy scores exceed 60% of the maximum observed score in the sentence are flagged as ambiguous and passed to Layer 2 as candidate assumption anchors. For cases meeting this threshold, the approach identifies candidate anchors without additional training and can be applied across requirement domains, unlike keyword-based methods that rely on static lexicons of vague terms.

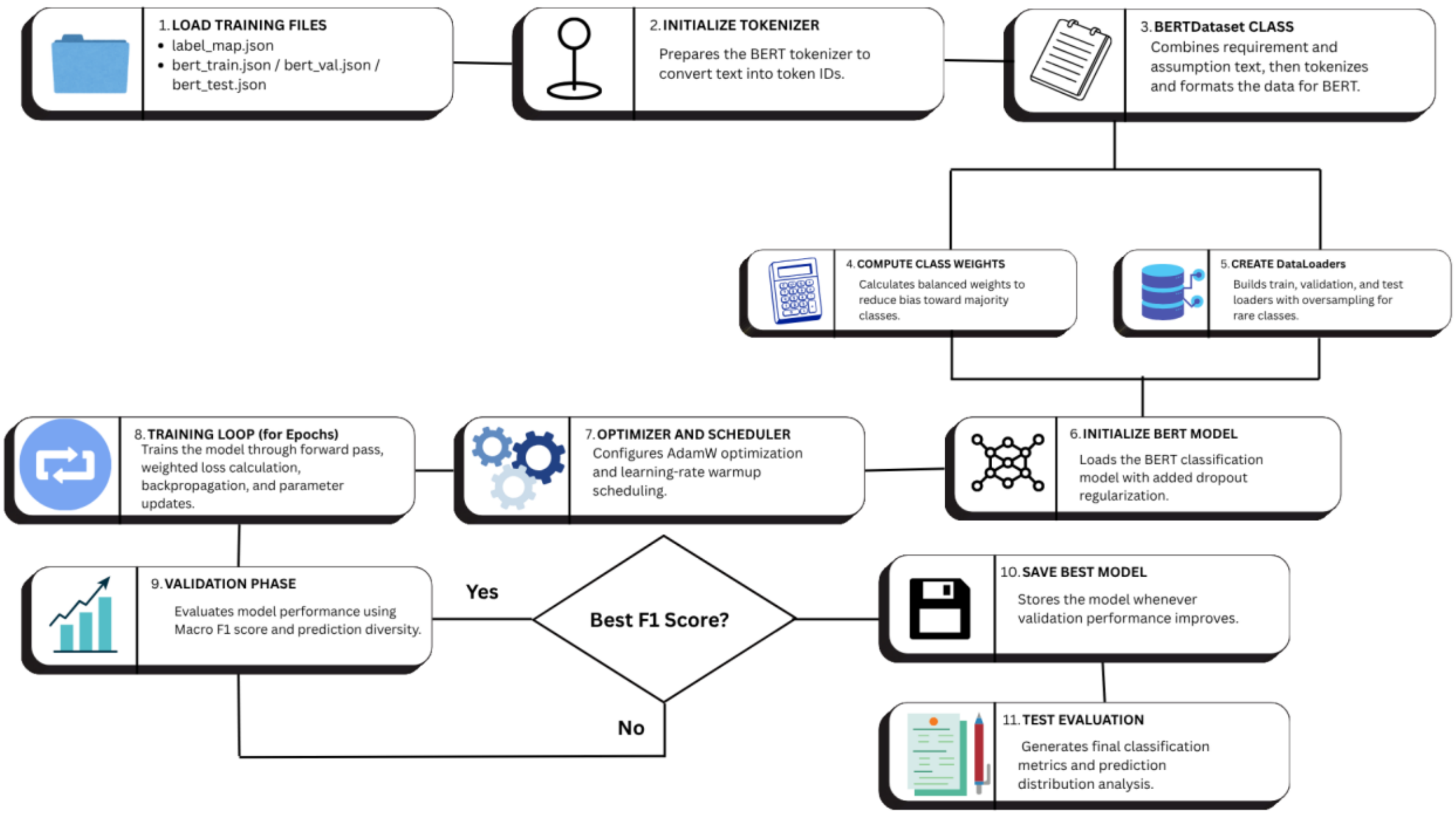

Figure 4. Layer 1 Architecture

Figure 4 depicts the workflow of Layer 1 visually. Firstly, the training files are loaded and the tokenizer is initialized to convert text into token IDs. Then the BERT Dataset CLASS combines requirements and assumptions in order for the class weights to be calculated. The dataloaders then build train, validation and test splits, which are then sent to the BERT model after it is initialized. Afterwards, the optimizer and scheduler configure AdamW optimization and figure out the scheduling of tasks. The training loop then takes over to train the model through forward pass, backpropagation, etc. The validation phase evaluates model performance for the best F1 score and saves the best model. Finally, the classification metrics are generated for review.

## 4.2 Layer 2: LLM Chain-of-Thought Assumption Detection

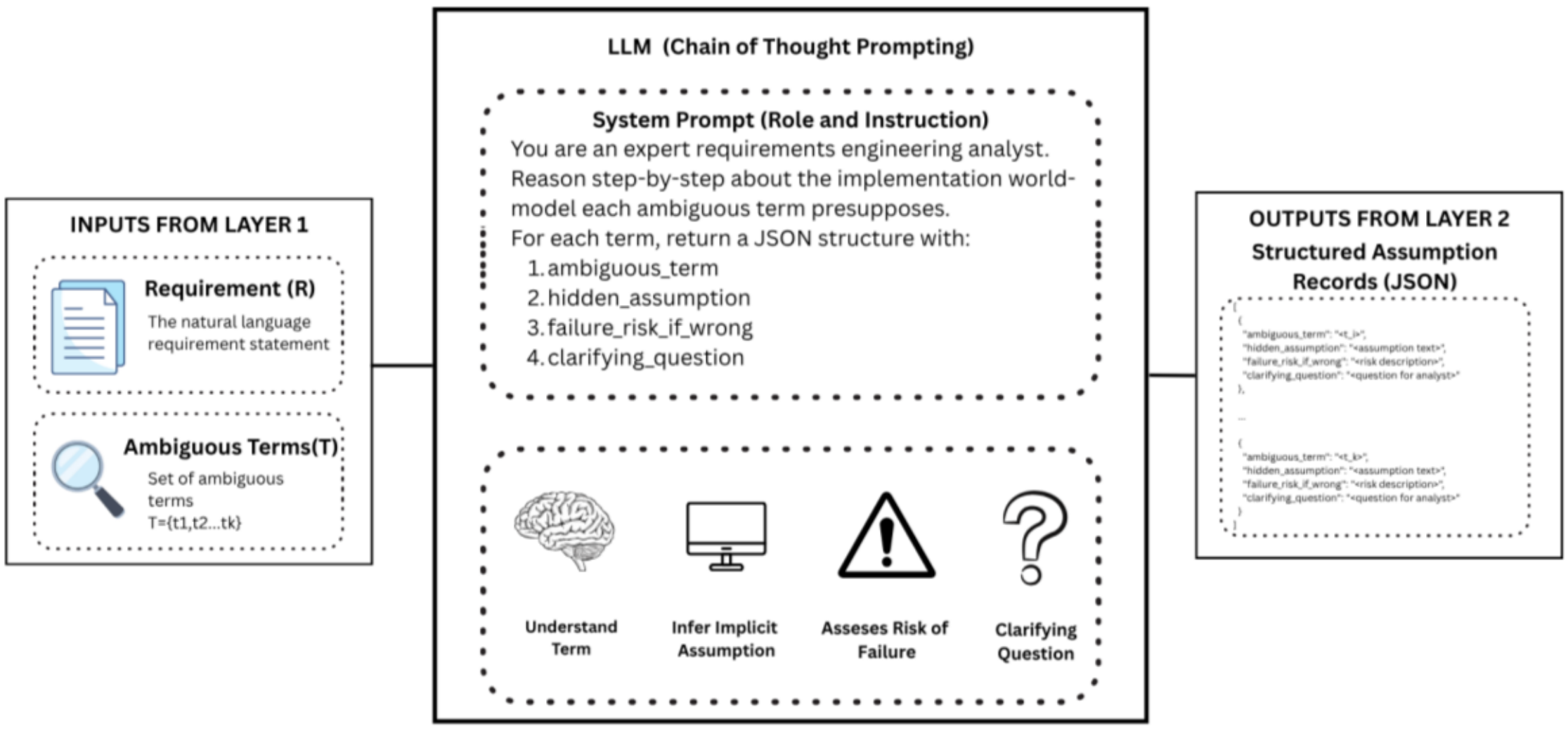

Figure 5. Layer 2 Architecture

Given the requirement $R$ and the set of ambiguous terms $T = \{t_1, \ldots, t_k\}$ from Layer 1, Layer 2 invokes an LLM with a CoT prompt to generate structured assumption records. The system prompt instructs the model to act as an expert requirements engineering analyst, reason step-by-step about the implementation world-model each ambiguous term presupposes, and return a JSON structure containing: (i) the ambiguous term, (ii) the hidden assumption text, (iii) the failure risk if the assumption is wrong, and (iv) a clarifying question for the analyst, all three components are shown pictorially through Figure 5 for better understanding.

### 4.2.1 Prompt Engineering Strategy

We evaluated three prompting strategies in Layer 2: Zero-shot (ZS) (Brown et al., 2020), few-shot (FS) (Brown et al., 2020), and CoT (Wei et al., 2022). In ZS mode, the model receives only the task description and the requirement. In FS mode, three hand-crafted requirement-assumption examples are prepended to the prompt. In CoT mode, few-shot examples include explicit reasoning traces that model the step-wise inferential process from requirement text to assumption. This progression mirrors the comparative methodology of Binkhonain and Alfayez (2025), who found that few-shot and persona-augmented CoT prompting matched or exceeded supervised baselines in requirements classification. Table 3 presents the strengths and weaknesses of the prompt strategy techniques used, and Figure 6 depicts the same via a detailed visual architecture.

**Table 3: Prompt Strategy Comparison**

| Strategy | Description | Strengths | Weaknesses |
|---|---|---|---|
| Zero-Shot (ZS) | Task description + requirement only. No examples. | No example overhead; fast. | Lower precision on implicit inference tasks. |
| Few-Shot (FS) | 3 labelled requirement-assumption pairs prepended. | Grounds model in task format; improves recall. | Sensitivity to example selection quality. |
| CoT | Few-shot examples augmented with step-by-step reasoning traces. | Best precision and coherence; structured JSON output. | Higher token cost; occasional over-explanation. |

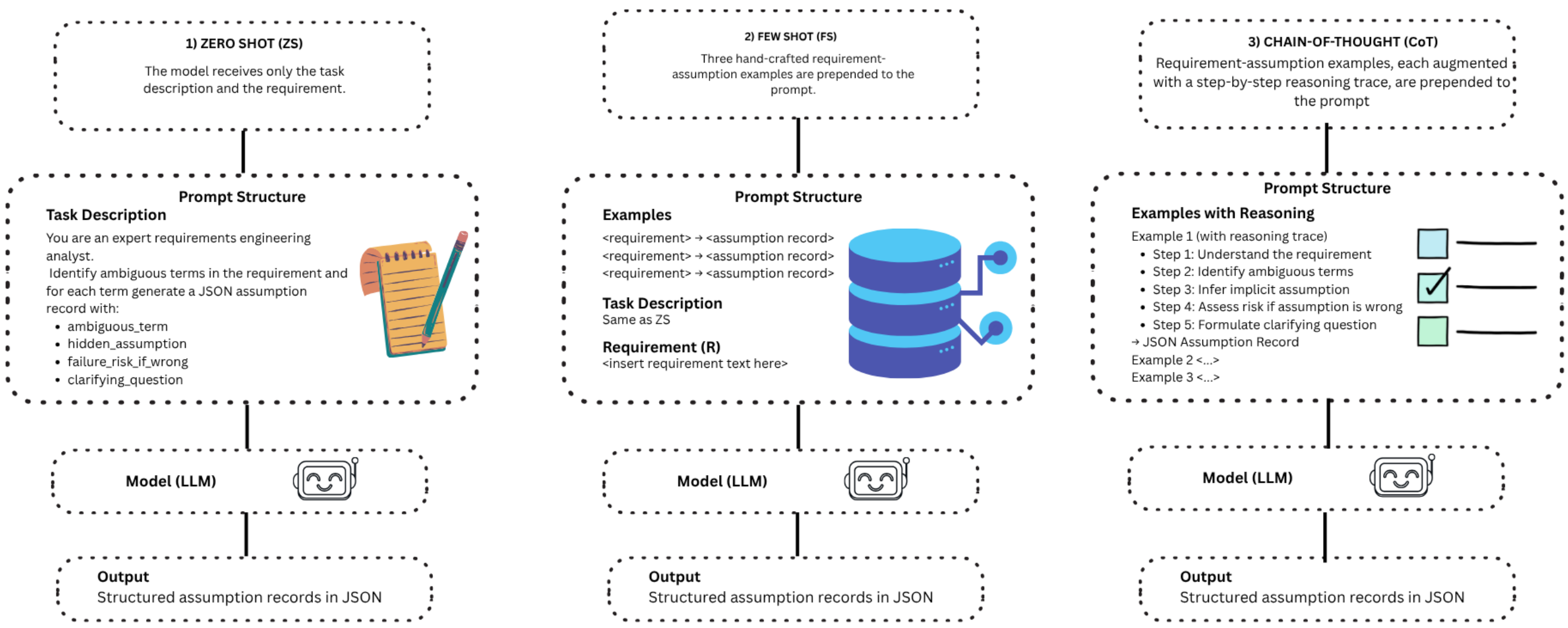

Figure 6. Prompt Engineering Strategy Architecture

Based on empirical evaluation, CoT prompting achieves the best precision-recall balance and is adopted as the default strategy. The full CoT prompt template is reproduced below:

*System: You are an expert requirements engineering analyst. Detect hidden assumptions in the given software requirement. For each assumption: (1) identify the ambiguous term, (2) state the hidden assumption explicitly, (3) describe the failure risk, (4) formulate a clarifying question. Reason step-by-step before generating JSON output.*
*User: Requirement: {requirement} | Ambiguous terms: {terms} | Think step-by-step then return JSON: {"assumptions": [{"ambiguous_term": ..., "assumption_text": ..., "failure_risk": ..., "clarifying_question": ...}]}*

### 4.3 Layer 3: Fine-Tuned BERT NFR Classifier

Each assumption hypothesis generated by Layer 2 is classified into one of the six NFR categories using a fine-tuned BERT for sequence classification model. As highlighted in Figure 7, the classifier receives the concatenated sequence [REQ] {requirement} [ASSUMPTION] {assumption} as input, enabling it to make category decisions that are sensitive to both the requirement context and the assumption content. Training details are provided in Section 6.

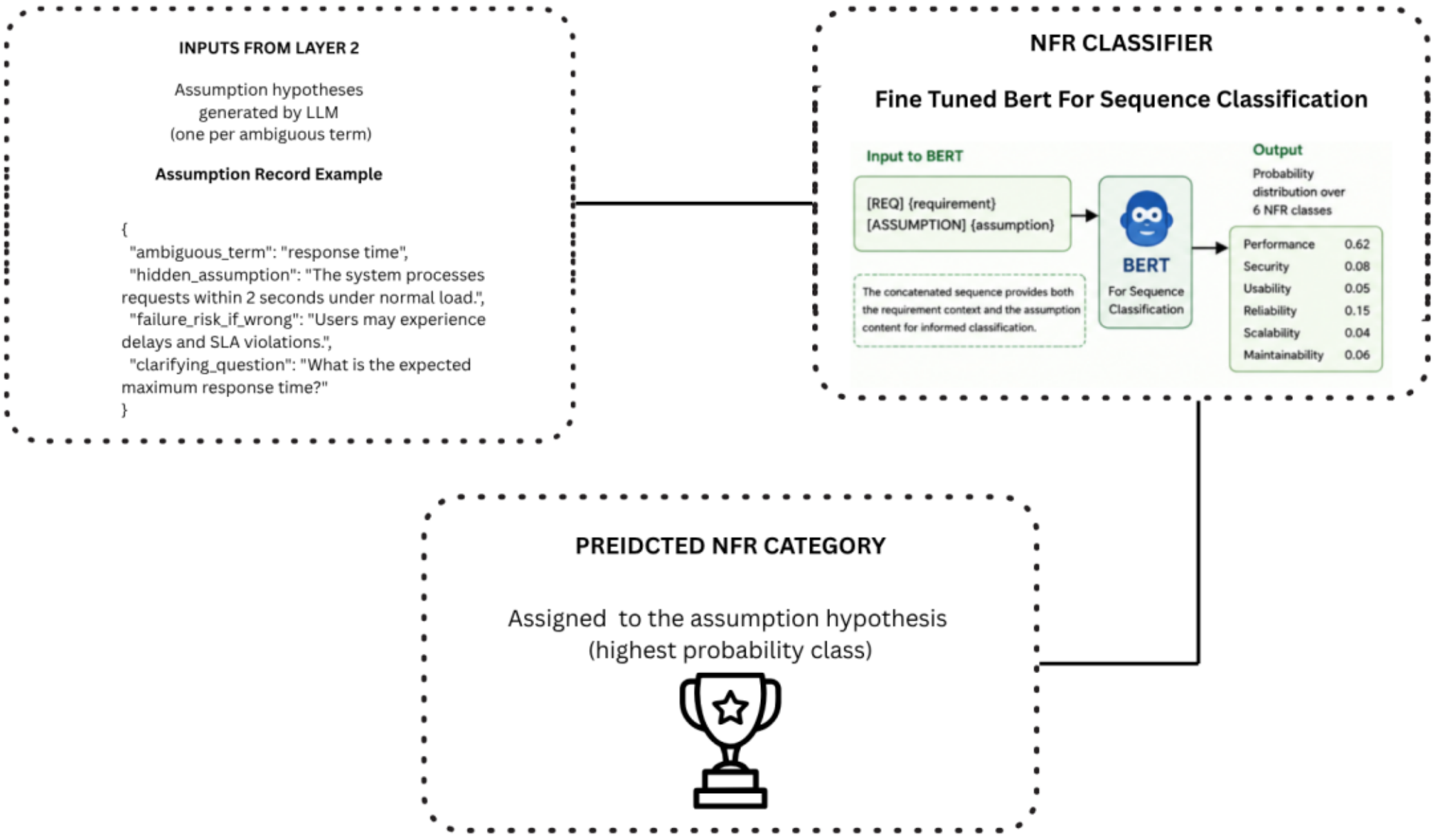

Figure 7. Layer 3 Architecture

### 4.4 Layer 4: LLM Clarification Suggestion Generation

Layer 4 generates a ranked list of three to five specific clarification options for the analyst for each classified assumption. The LLM is prompted with the requirement, the assumption text, the NFR category and the clarifying question, and asked to produce specific, actionable answer options that an analyst can choose or adapt. For instance, if we assume that '*quickly*' means a threshold of performance, the layer would produce options like '*within 200 ms*', '*within 500 ms for 95th percentile*', and '*within 1 second under peak load*'. A sub-function of layman mode rephrases the clarifying question in plain English for non-technical stakeholders. A visual flowchart that elaborates the same process in detail is shown in Figure 8.

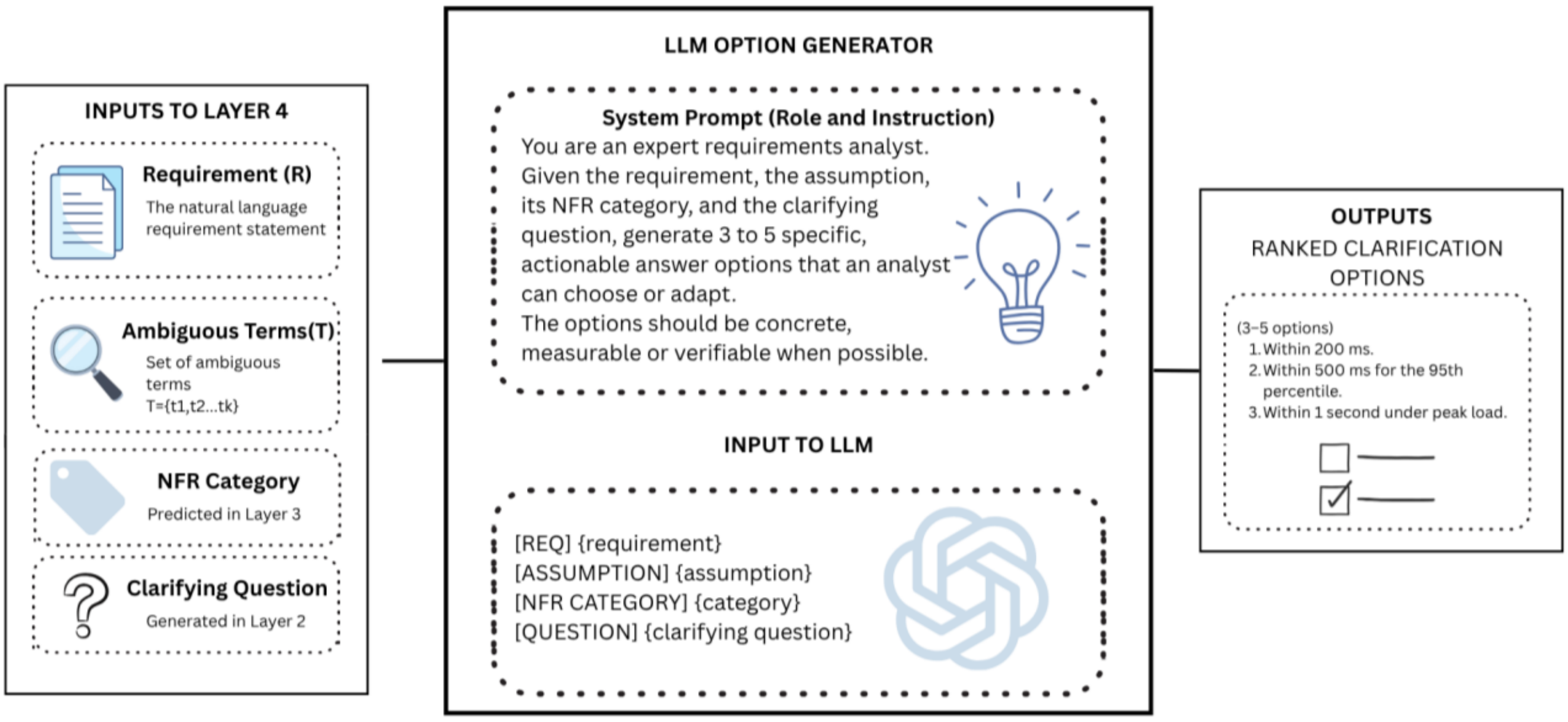

Figure 8. Layer 4 Architecture

## 4.5 Layer 5: Analyst Validation Loop

HiddenLens incorporates a human-in-the-loop validation step, where the analyst reviews each assumption, its categorisation and the suggested clarifications. The analyst can choose a pre-generated answer, type a custom free-text answer (validated for relevance and coherence by the LLM), or ask for a layman rephrasing of the clarifying question. This design is aligned with the results of Unterbusch and Vogelsang (2026), who showed that human-LLM collaborative validation substantially outperforms fully automated approaches on requirements quality tasks. Inputs to Layer 5 include Requirements, Ambiguous Terms, NFR category, Clarifying Question and Suggested Option, and all of these are subsequently passed through an Analyst Review Interface and Analyst Actions as shown in Figure 9.

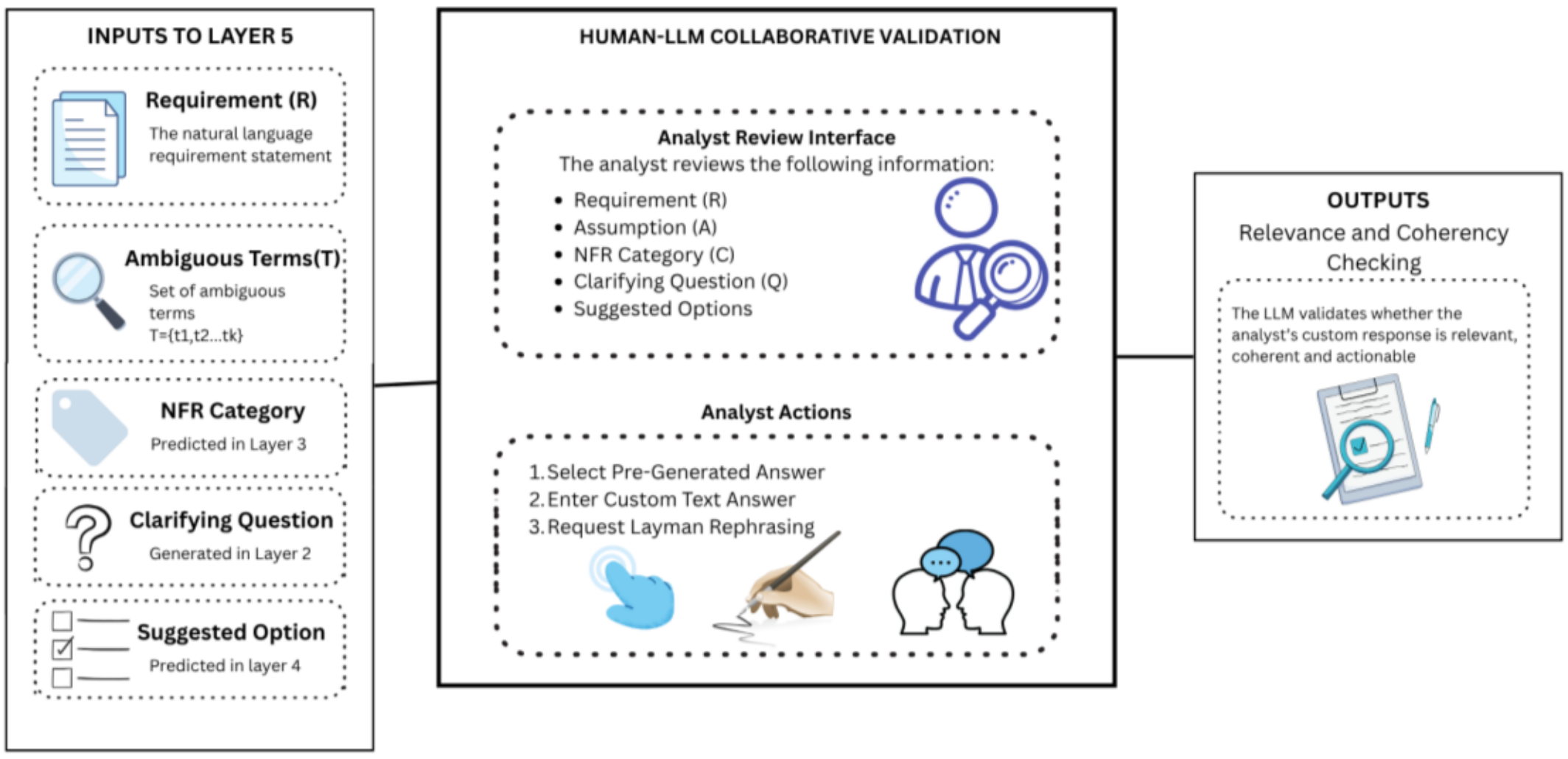

Figure 9. Layer 5 Architecture

## 4.6 Layer 6: Requirement Rewriter

Once all clarifications are collected, Layer 6 rewrites the original requirement by replacing each ambiguous term with the analyst-confirmed resolution using boundary-aware regular expression matching. The output is a refined requirement that explicitly states the assumptions that were implicit previously, as shown in Figure 10. For example, '*The system shall process user requests quickly*' would be '*The system shall process user requests within 200 ms at 95th percentile*' under 1,000 concurrent users. Algorithm 1 below depicts the flow of the hidden assumption detection pipeline followed in HiddenLens.

**Algorithm 1: HiddenLens Hidden Assumption Detection Pipeline**

```
Input:  R - natural language requirement sentence
Output: R' - rewritten requirement with explicit assumptions

T ← BERT_AttentionEntropy(R, top_n=4, threshold=0.60)  // Layer 1
if |T| = 0 then return R  // no ambiguous terms found
A ← LLM_CoT_Detect(R, T)  // Layer 2: assumption hypotheses
for each a ∈ A do
    a.category ← BERT_Classify([REQ] R [ASSUMPTION] a)  // Layer 3
    a.suggestions ← LLM_Suggest(R, a, a.category)  // Layer 4
end for
C ← {}  // clarification map
for each unique term t in T do
    c ← Analyst_ValidateLoop(t, A[t], R)  // Layer 5
    C[t] ← c
end for
R' ← Rewrite(R, C)  // Layer 6
return R'
```

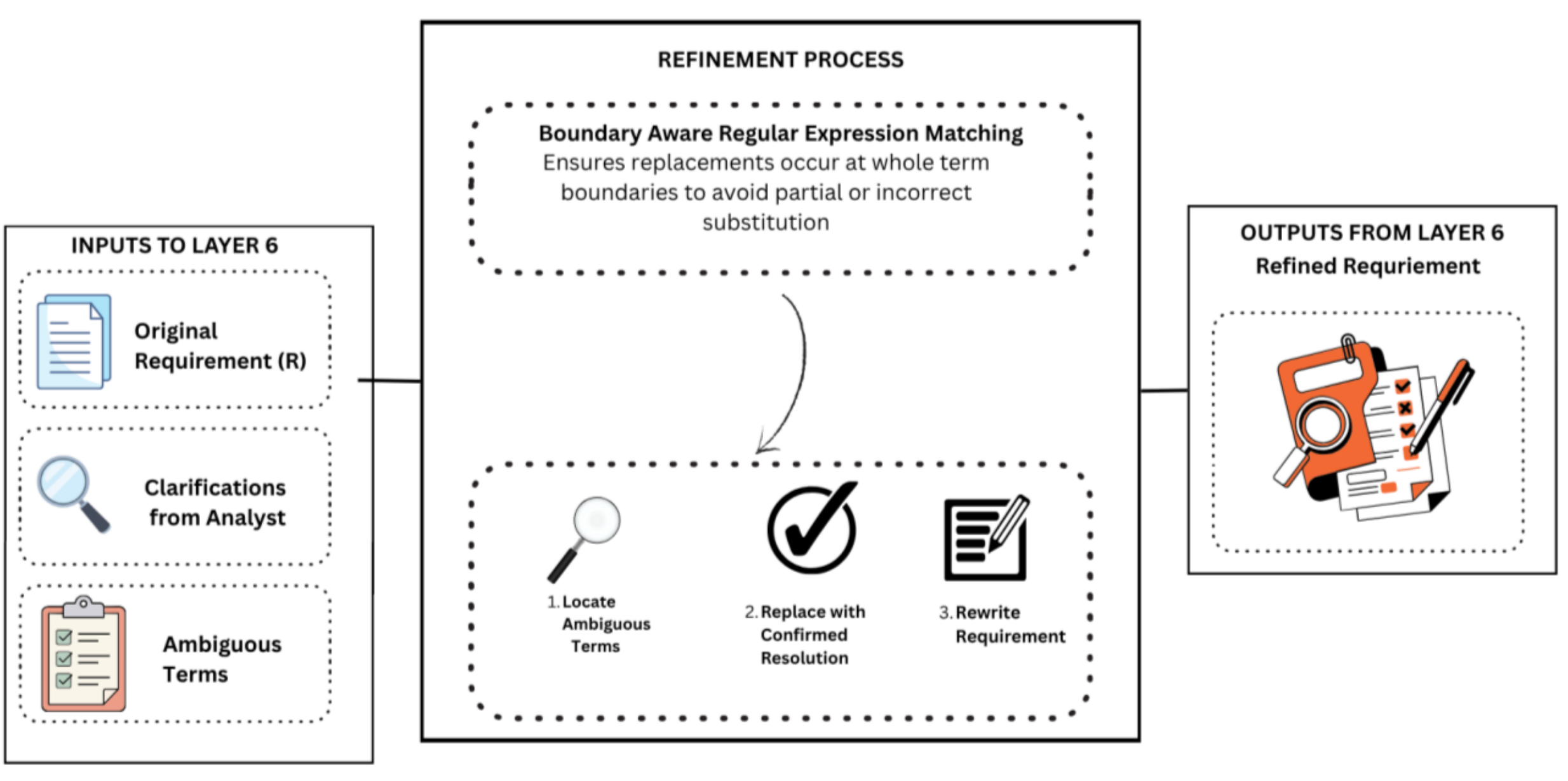

Figure 10. Layer 6 Architecture

The next section describes how the dataset used to train and evaluate HiddenLens was built, since no existing dataset of hidden assumptions in software requirements was available for this purpose.

## 5. Dataset and Annotation

Since no annotated corpus dealing with hidden assumptions in software requirements currently exists, we have created a dedicated dataset for this study: the **HiddenLens Corpus (HLC)**, addressing what Abualhaija et al. (2024) identify as a systemic barrier in NLP-for-RE research, where annotated datasets remain scarce, and only a small fraction of published tools or data are ever made publicly available. The requirements were obtained from different sources, namely: SRS documents publicly available in academic repositories such as the PROMISE NFR dataset (Cleland-Huang et al., 2007) and PURE dataset (Ferrari et al., 2017) and requirements from the novel HiddenLens Corpus (HLC). The final dataset, HLC, comprises 312 requirements containing 847 annotated assumptions

The workflow is illustrated in Figure 11. It begins by randomly shuffling the list of requirements to minimise category-ordering bias. The code then iterates through each requirement and first determines whether all six NFR categories—performance, security, scalability, usability, reliability, and domain—have reached the maximum limit of 60 items per category. If this condition is satisfied, the loop terminates. For each eligible requirement, the *annotate_requirement( )* function sends the requirement text to the language model through the OpenRouter API. The temperature is set to 0.3, and the request is retried up to three times in the event of a rate-limit error. The result is stripped of Markdown fences and parsed as JSON to get the inferred NFR category and list of assumptions. If there are no assumptions in the result, or if the inferred category is full, this item is simply skipped. Otherwise, the result is added to the output list, and the respective category counter is increased by one. Progress and distribution by categories are logged every ten items. Sleeping for half a second after each iteration ensures that the API rate limit is not exceeded. Once the loop completes, the saved items from previous runs are loaded and added to the new annotated results. The final dataset is saved to a file, and a summary is printed.

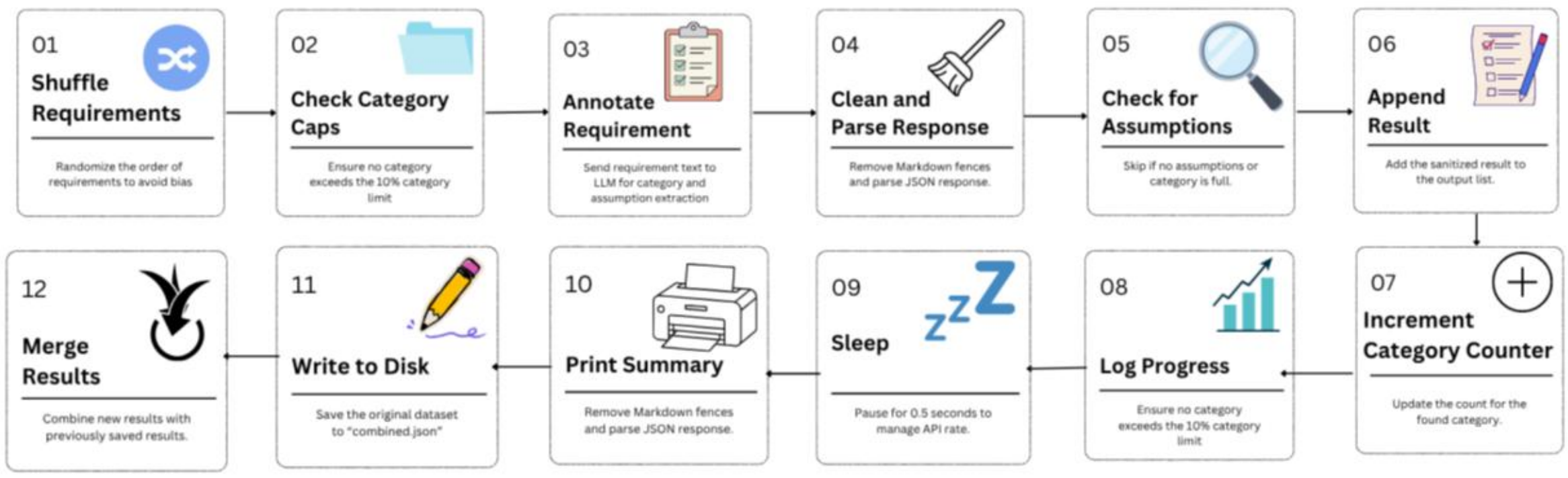

Figure 11. Automated Requirement Annotation Process

The dataset was partitioned into training (70%), validation (15%), and test (15%) splits using stratified sampling to preserve category proportions, as detailed in Table 4. For BERT classifier training (Layer 3), samples are requirement-assumption pairs labelled with the assumption category. For T5 generative model training (Layer 2 offline fallback), samples are (requirement, structured assumption string) pairs.

**Table 4: HiddenLens Corpus (HLC) Statistics**

| Category | Requirements | Assumptions |
|---|---|---|
| Performance | 52 | 148 |
| Security | 61 | 174 |
| Scalability | 44 | 118 |
| Usability | 48 | 127 |
| Reliability | 59 | 163 |
| Domain | 48 | 117 |
| TOTAL | 312 | 847 |

# 6. Experimental Setup

This section describes the configuration of each model used in HiddenLens, together with the baseline methods against which the framework is evaluated.

## 6.1 Model Configuration

Table 5 provides an overview of the parameters used across the three components of HiddenLens, namely, the BERT classifier in Layer 3, the T5 fallback generator in Layer 2 offline mode, and the LLM used for CoT-based detection of assumptions and generation in Layers 2 and 4.

**Table 5:** Model Configuration Parameters used in HiddenLens

| Component | Parameter | Value |
|---|---|---|
| BERT Classifier (L3) | Base model | bert-base-uncased |
| | Max sequence length | 256 tokens |
| | Batch size | 16 |
| | Learning rate | 2e-5 |
| | Epochs | 5 (best validation F1) |
| T5 Generator (L2 fallback) | Base model | t5-small |
| | Batch size | 8 |
| | Learning rate | 3e-4 |
| LLM (Layers 2, 4) | Model | meta-llama/llama-3.3-70b-instruct |
| | Temperature | 0.2 (detection), 0.7 (suggestions) |
| | Max tokens | 1000 |
| | Prompt strategy | Chain-of-thought (default) |

BERT-base-uncased was chosen as the base model for NFR classifier because of its excellent performance in sequence classification tasks, having been extensively studied in terms of training dynamics, and being easy to fine-tune owing to the availability of computational hardware. It has been noticed that 256 tokens is the optimal token length for a sequence, as fewer than 0.3% of the input sequences required truncation. The learning rate of 2e-5 with early stopping on validation F1 follows the fine-tuning recommendations established for BERT-family models. For the LLM component, the temperature of 0.2 for assumption detection reflects the requirement for deterministic, factual inference: higher temperatures introduced variability in the logical structure of assumption statements that reduced inter-run consistency. The elevated temperature of 0.7 for suggestion generation is intentional, this task benefits from lexical diversity in the generated clarification options, giving analysts a richer menu of choices rather than semantically near-identical variants of the same resolution.

## 6.2 Baseline Methods

We compare HiddenLens against three baseline methods. (1) Keyword-Based (KB): A rule-based system (Wilson et al., 1997) that flags requirements containing vague terms from a curated lexicon (e.g., '*quickly*', '*efficiently*', '*securely*', '*in real time*') and retrieves pre-written generic assumption statements for each term. (2) QuARS-Adapted (QA) (Lami et al. 2005): The QuARS syntactic analyser extended with assumption-class labels assigned to implicit, under-specification, and vagueness detections. (3) Zero-Shot LLM (ZS-LLM) (Brown et al., 2020): The Layer 2 LLM component with zero-shot prompting only, without BERT attention pre-filtering, CoT reasoning traces, or few-shot examples. This baseline isolates the contribution of our prompt engineering and architectural integration choices.

## 6.3 Evaluation Metrics

HiddenLens is put to the test using two methodologies, namely: automated metrics against held-out annotations and human evaluation by RE practitioners.

### 6.3.1 Automated Metrics

Macro-averaged Precision, Recall and F1 on the test split for Layer 3's BERT classifier are reported as follows:

$$P = TP / (TP + FP), \quad R = TP / (TP + FN), \quad F1 = 2 \times P \times R / (P + R)$$

A predicted assumption is counted as a true positive in Layer 2 if the text has a ROUGE-L overlap ≥ 0.4 with any gold annotation for the same requirement, reflecting semantic rather than verbatim match.

### 6.3.2 Human Evaluation

The test split gave rise to requirement-assumption pairs and was presented to humans who did not participate in the annotation. Each practitioner rated the assumptions on three criteria using a 5-point scale as depicted in Table 6.

**Table 6: Human Evaluation Criteria**

| Criterion | Definition | Rating Scale |
|---|---|---|
| Relevance | Does the assumption represent a genuine implicit belief that affects implementation? | 1 (irrelevant), 5 (highly relevant) |
| Clarity | Is the assumption stated in clear, actionable language that an analyst can act upon? | 1 (unclear), 5 (very clear) |
| Usefulness | Would addressing this assumption improve requirement quality and reduce implementation risk? | 1 (not useful), 5 (very useful) |

## 7. Results and Discussion

This section presents a quantitative evaluation of HiddenLens against the baseline models and previous approaches, followed by representative outputs illustrating the framework's performance in practice.

### 7.1 Quantitative Results

HiddenLens with CoT prompting achieves an F1 of 0.79 on assumption detection, representing a 71% relative improvement over the keyword-based baseline and a 61% improvement over QuARS-Adapted (Lami et al. 2005). The comparison between ZS-LLM and HiddenLens ZS demonstrates that BERT attention pre-filtering alone contributes approximately 3 F1 points by constraining the LLM search space to genuine ambiguous terms. The progression from zero-shot through few-shot to CoT confirms the findings of Unterbusch and Vogelsang (2026) that reasoning traces substantially improve LLM performance on implicit inference tasks. All results are presented in a tabular format below through Table 7, and further visual depiction is presented in Figure 12.

**Table 7: Performance Results- Assumption Detection**

| Method | Precision | Recall | F1 |
|---|---|---|---|
| Keyword-Based (KB) | 0.41 | 0.53 | 0.46 |
| QuARS-Adapted (QA) | 0.47 | 0.51 | 0.49 |
| Zero-Shot LLM (ZS-LLM) | 0.63 | 0.61 | 0.62 |
| HiddenLens (ZS prompt) | 0.66 | 0.64 | 0.65 |
| HiddenLens (FS prompt) | 0.73 | 0.72 | 0.72 |
| HiddenLens (CoT prompt) | **0.81** | **0.78** | **0.79** |

The BERT NFR classifier (Layer 3) achieves macro F1 Score of 0.847 on the held-out test split, with highest performance on Security (F1 Score 0.89) and lowest on Domain (F1Score 0.79), reflecting the inherent breadth and context-dependence of the Domain category. These results are competitive with NoRBERT (Hey et al., 2020) on the PROMISE NFR benchmark PROMISE (Cleland-Huang et al., 2007), despite our classifier addressing a novel assumption-classification task rather than standard requirement type classification.

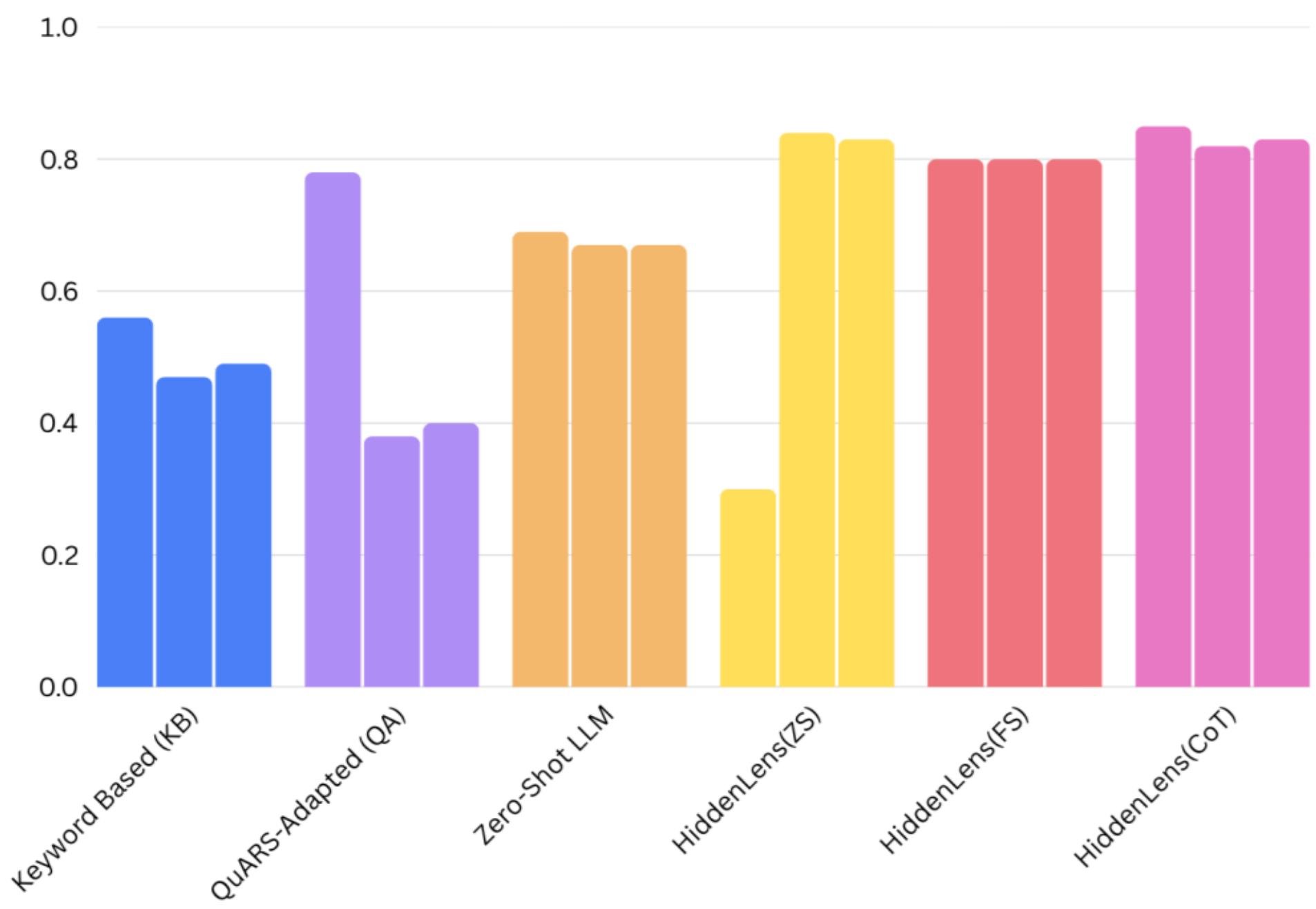

Figure 12. Performance comparison (Precision, Recall and F1) between HiddenLens and other baseline models

### 7.2 Comparative Analysis with Prior Work

Each cited work is a real contribution to its target problem (requirements smell detection, NFR classification, defect prediction), but none of them were designed with implicit content as its main object of analysis. The closest prior work to HiddenLens is by Femmer et al. (2017), which considers requirements quality as a property of the text itself, rather than a mapping to an external standard. The most methodologically similar work is the work of Unterbusch and Vogelsang (2026) on defect prediction, which is also committed to CoT prompting and human-LLM collaboration like HiddenLens. More such comparisons are presented in Table 8. The focus on defective explicit content and not implicit omissions means that the architecture and training data are not directly transferable, but the empirical findings are a strong validation for the methodological choices taken by HiddenLens. It builds on that foundation by addressing the assumption gap that remains one of the most significant unresolved problems in the field.

**Table 8: Comparative Analysis with Prior Approaches of Detecting hidden assumptions in RE**

| Authors/Year | Technique | Task | Limitation | Our Proposed Work - HiddenLens |
|---|---|---|---|---|
| Femmer et al. (2017) | POS rules + heuristics | Requirements smell detection | No implicit content | ✓ |
| Hey et al. (2020) | BERT transfer learning | NFR classification | Typed text only | ✓ |
| Luo et al. (2022) | Prompt BERT (PRCBERT) | Zero-shot RE classification | No assumption output | ✓ |
| Unterbusch and Vogelsang (2026) | CoT + human-in-loop | Defect prediction | Defects, not assumptions | ✓ |
| Ferrari et al. (2024) | LLM ambiguity detection | Ambiguity classification | Stated content only | ✓ |
| Zadenoori et al. (2025) | SLR of LLMs in RE | Survey | No tool; no assumptions | ✓ |

### 7.3 Input and Output - HiddenLens

Table 9 presents three representative examples of HiddenLens's Layer 2 and Layer 4 output, spanning three different NFR categories: reliability, performance, and security. Each row shows a single vague term extracted from a real requirement, the specific hidden assumption HiddenLens infers behind it, and the clarifying question generated for the analyst to resolve it, illustrating how the framework moves from an ambiguous word to a concrete, answerable specification gap.

**Table 9: HidenLens Input and Output**

| Requirement | Detected Assumption | Generated Question |
|---|---|---|
| The system shall respond immediately in emergencies. | [Reliability] "Immediately" implies sub-second alert delivery to on-call mobile devices, assuming reliable network availability. | What is the maximum acceptable response time in an emergency scenario, and what channels must receive the alert? |
| Doctors can update patient records quickly. | [Performance] "Quickly" implies record update operations complete in < 500 ms at 95th percentile under concurrent clinical load. | What response time threshold defines "quickly" for record updates, and under what user concurrency? |
| Data shall be stored securely. | [Security] "Securely" implies AES-256 encryption at rest, TLS 1.3 in transit, and RBAC with least-privilege access. | What encryption standard and access control model is required, and which regulatory framework applies? |

## 8. Case Study – Hospital Patient Management System

This section presents the HiddenLens architecture and demonstrates its six-layer pipeline through a case study involving a *Hospital Patient Management System*. The requirement provided by the stakeholders in natural language is as follows:

*"The Hospital Patient Management System allows doctors, nurses, and administrators to manage patient records efficiently. Doctors can view and update patient records quickly. Nurses can administer medication and update patient status in real time. The system must ensure patient data is stored securely and only accessible to authorized personnel. In emergency situations, the system should respond immediately and notify the relevant staff. Administrators generate daily reports and manage staff schedules regularly."*

Each of the six layers of HiddenLens pipeline was applied sequentially to these requirements. In layer 1, vague terms such as quickly, real time, securely, authorized personnel, immediately, and regularly were identified. Assumption hypotheses for each of the above terms were created for layer 2. For instance, assumption hypotheses that implied two seconds response time assumed by the developer were created for the term quickly, and assumption hypotheses based on role-based access control and at-rest encryption were made for the term securely. These assumptions were then categorized using a BERT classifier into NFR categories (performance, security, scalability, and domain compliance). Then, in Layer 4, LLM-driven clarification suggestions were provided with chain-of-thought prompting, for instance, replacing the phrase “Doctors can view and update patient records quickly” with “The system shall display a patient record within 200 ms under normal load and within 500 ms under peak load.” After validation from the analyst in Layer 5, Layer 6 performed regex-based substitution for rewriting the requirements.

In order to illustrate the improvement in requirement clarity that HiddenLens makes possible, consider the case of the Hospital Patient Management System requirements before and after processing. There are five requirements that use six fuzzy terms in the process: "quickly," "real time," "stored securely," "authorized personnel," "immediately," and "regularly." All of these terms lack measures, scopes, and testable criteria for acceptance. If a software developer were to read these requirements, there would be no common point from which verification could be done, and the developer would be left to determine his/her own criteria for each term.

After passing through all six layers of HiddenLens, all requirements are tied to concrete, verifiable specifications: response times are specified in milliseconds with load conditions appended, encryption is linked to named standards (AES-256, TLS 1.3), access control is linked to a documented Role-Based Access Control (RBAC) policy with mandatory audit logging, emergency notifications come with a three-second Service-Level Agreement (SLA) with named recipient roles, and administrative scheduling is scoped to a defined staff ceiling with a conflict-resolution deadline. The change is from requirements that are syntactically correct but semantically underspecified, that is, requirements that might pass a casual review but would fail at implementation or acceptance testing, to the requirements that meet the ISO/IEC/IEEE 29148 (ISO/IEC/IEEE, 2018) criteria of unambiguity, verifiability and completeness. This improvement was made without losing the original intent of any requirement. HiddenLens surfaces and resolves hidden assumptions rather than replacing stakeholder decisions, preserving the what while making the how measurable.

Table 10 below details the output in a tabular format showcasing a 'before vs after' of the SRS documentation. In the rewritten SRS, the IEEE standard followed has also been depicted for better viability. Figure 13 presents the output displayed in the Google Colab environment, whereas Figure 14 shows a representative step in the rewriting process at which the user is prompted to select an option.

Table 10. SRS Case Study before HiddenLens rewrite vs. after rewrite

| **Req. ID** | **NFR Category** | **Before HiddenLens: Original Requirement** | **Ambiguous Terms** | **Hidden Assumption** |
|---|---|---|---|---|
| **R1** | *Performance* | Doctors can view and update patient records **quickly**. | Quickly | Developer assumes <2s response; no load condition stated. |
| **R2** | *Performance / Scalability* | Nurses can administer medication and update patient status in **real time**. | real time | Implies instant sync; propagation delay and concurrency undefined. |
| **R3** | *Security / Domain* | The system must ensure patient data is **stored securely** and only accessible to **authorized personnel**. | stored securely, authorized personnel | Encryption standard, RBAC policy, and audit logging all unstated. |
| **R4** | *Performance / Domain* | In emergency situations, the system should respond **immediately** and notify the **relevant staff**. | immediately, relevant staff | Threshold undefined; notified roles and channels unspecified. |
| **R5** | *Scalability / Domain* | Administrators generate daily reports and manage staff schedules **regularly**. | Regularly | Schedule frequency, staff volume, and conflict resolution SLA all implicit. |
| **After HiddenLens: Rewritten Requirements** | | | | |
| **Req. ID** | **NFR Category** | **After Hidden Lens Rewrite** | **Resolved Specifications** | **ISO/IEC/IEEE 29148 Criteria Met** |
| **R1** | *Performance, Scalability* | The system shall display a patient record within **200 ms** under normal load and within **500 ms** under peak load (≤**500 concurrent users**). | 200 ms 500 ms 500 concurrent users | Verifiable, Unambiguous |
| **R2** | *Performance, Scalability* | Medication administration updates shall be reflected system-wide within **2 seconds** of submission, with a maximum propagation delay of **5 seconds** under peak ward load. | 2 seconds 5 seconds | Verifiable, Complete |
| **R3** | *Security, Domain Compliance* | Patient data shall be encrypted at rest using **AES-256** and in transit using **TLS 1.3**. Access shall be restricted to roles defined in the **RBAC policy**; all access events shall be audit-logged. | AES-256 TLS 1.3 RBAC policy | Verifiable, Unambiguous, Complete |
| **R4** | *Performance, Domain Compliance* | Emergency alerts shall be dispatched to on-call personnel within **3 seconds** of trigger. At minimum the **attending physician** and **ward nurse** roles shall receive notification via pager and in-app alert. | 3 seconds attending physician ward nurse | Verifiable, Unambiguous |
| **R5** | *Scalability, Domain Compliance* | Administrators shall generate reports on a **configurable schedule** (minimum: daily). The scheduling module shall support up to **200 staff members** and resolve conflicts within **one working day**. | configurable schedule 200 staff members one working day | Verifiable, Complete |

```
Ambiguous term : "reports"  [Performance]
Failure risk   : System may not provide the required reports or may not support ad-hoc reporting.
Question       : What specific types of reports are required and what data should they contain?
────────────────────────────────────────────────────────────────
Suggested options (LLM-generated, context-aware):
  [1] summary logs of system usage and errors
  [2] detailed analytics of staff productivity metrics
  [3] compliance audit trails with timestamped events
  [4] key performance indicator dashboards with real-time data
  [C] Enter a custom answer
  [L] Show plain-English / layman version of this question

Your choice: 3
  -> Selected: "compliance audit trails with timestamped events"

======================================================================
REWRITTEN REQUIREMENT
Administrators produce in Excel format on demand every weekday morning by 8am compliance audit trails with timestamped events and update at least 95% of staff schedule
staff schedules regularly.
======================================================================

======================================================================
FINAL REWRITTEN PARAGRAPH
The Hospital Patient retrieval, updating, and archiving of patient data with 99.9% accuracy System supports a minimum of 100 concurrent users healthcare professionals
to create, update, and delete patient records efficiently. Doctors can They should be able to see the patient's medical history, current medications, and test results.
in under 1 second for 95% of user interactions. Nurses can inject or orally provide medication within 2 minutes of electronic order receipt medication and commit updat
authorized personnel. In emergency situations, the system should generate a response message within 500 milliseconds within 1 second
and trigger a audible alarm and flash the users with administrative or supervisory roles member's workstation screen for 10 seconds the relevant staff. Administrators
staff schedules regularly.
======================================================================
```

Fig 13. Rewritten Case Study Output

```
────────────────────────────────────────────────────────────────
Suggested options (LLM-generated, context-aware):
  [1] display patient records within 2 seconds
  [2] render patient records in a readable format
  [3] access patient records with a single mouse click
  [4] retrieve and display patient records in under 1 second
  [C] Enter a custom answer
  [L] Show plain-English / layman version of this question

Your choice: L
  [Fetching plain-English version from OpenRouter...]

  Plain-English question:
  >> What exactly do you want doctors to be able to see or do when they look at a patient's record?
  Example good answer  : They should be able to see the patient's medical history, current medications, and test results.
```

Fig 14. Output Screen

## 9. Threats to Validity

This section discusses the threats to the validity of our findings, organized across internal, external, construct, and LLM-specific dimensions.

**Internal validity:** The main risk arises from the subjective nature of annotations. Although the inter-annotator agreement ($\kappa = 0.74$ on assumption presence) is high, the difference between hidden assumptions and implicit expectations of stakeholders is not always clear-cut. Formal definitions of Section 3.1 reduce but do not entirely resolve this issue. The systematic annotator bias could over- or underestimate some categories of NFRs.

**External validity:** The findings are based on a dataset compiled from three specific sources. Norheim et al. (2024) observed that published requirements datasets, which typically contain between a few dozen and a few thousand requirements, remain substantially smaller than real-world industrial specifications, which may contain hundreds of thousands of requirements. The 312-requirement HLC is subject to the same limitation. Consequently, the transferability of the findings to assumptions in requirements from other domains, such as embedded systems, telecommunications, and intelligent systems, requires domain-specific validation. Furthermore, industrial requirements may be more concise than examples reported in the scholarly literature, potentially resulting in a higher density of assumptions per sentence.

**Construct validity:** The chosen criterion for identifying TPs, ROUGE-L overlap $\geq$ 0.4 is an acceptable approximation only. Gold standard assumptions that employ specialized domain-specific terminology not contained in the text of the predicted assumption will be automatically counted as False Negatives (FN). The gold assumptions, which incorporate terms unique to the field that are not included in the text of the assumed hypothesis, can mistakenly be considered false negatives even with identical semantics.

**Limitations of LLM:** It is possible for LLMs to generate assumptions that appear logical but are not really necessary for the implementation process. Layer 5, which is the validation loop, is the safety valve but leaves the responsibility on the shoulders of the analyst.

## 10. Conclusion and Future Work

This paper proposes **HiddenLens**, a novel six-layer BERT-LLM pipeline and the first framework engineered specifically to detect, classify, question, and resolve hidden assumptions in natural language software requirements. The framework leverages BERT attention entropy with LLM reasoning, BERT fine-tuning for NFR classification, context-aware suggestion generation, and human-in-the-loop validation to address an aspect of requirements quality that has hitherto received inadequate attention: the hidden world-view beliefs about the nature of the world that drive design choices in implementations.

Through experimental results on the 312 requirements in the HiddenLens Corpus (HLC), annotated with hidden assumptions, we achieve an F1 score of 0.79 for assumption identification and an F1 score of 0.847 for BERT classifier-based NFR classification. The case study on requirements from a Hospital Management System is one such illustration of the real-world importance of HiddenLens for safety-critical systems. To the best of our knowledge, **HiddenLens** represents the first dedicated attempt to address the detection and resolution of hidden assumptions in requirements engineering (RE). It can assist analysts, developers, and organisations in identifying potential problems before they become costly or contribute to late-stage project failures, particularly when such problems cannot be identified by conventional ambiguity-detection tools during the early stages of the software development life cycle (SDLC).

Future directions include incorporating HiddenLens into commercial software development tools like IDEs and requirement management tools in order to investigate the influence of HiddenLens on projects over time. Overall, HiddenLens aims to close the gap between what a requirement states and what it silently assumes, turning tacit engineering judgment into an explicit, testable artifact.

**Data Availability Statement**
The data (HiddenLens Corpus) and codes developed and analysed during the current study will be made publicly available in a Github repository upon acceptance of the manuscript.

**Declaration of generative AI and AI-assisted technologies in the manuscript preparation process**
During the preparation of this manuscript, the author(s) used ChatGPT-5.5 (OpenAI) to improve sentence structure in selected passages. The author(s) reviewed and edited the output as needed and take full responsibility for the content of the published article.